\documentclass[11pt,a4paper]{article}
\pdfoutput=1
 
\usepackage[nosort]{cite}

\usepackage[british]{babel} 

\usepackage{epsfig,rotating}
\usepackage{amsmath,amssymb,amsthm}
\usepackage{amsfonts}
\usepackage{mathrsfs}
\usepackage{bbm}
\usepackage{bm}
\usepackage[normalem]{ulem}
\usepackage[all,knot]{xy}
\xyoption{arc}

\usepackage{tikz}
\usetikzlibrary{arrows}
\usetikzlibrary{patterns}
\usetikzlibrary{decorations.markings}

\usepackage{mathabx}

\usepackage{xcolor}
\usepackage[textwidth = 430 pt, textheight = 630 pt]{geometry}
\definecolor{MyDarkBlue}{rgb}{0.15,0.25,0.45}
 
\usepackage{enumerate}
 
\usepackage[linktocpage=true,hypertexnames=false]{hyperref}
\hypersetup{
colorlinks=true,
citecolor=MyDarkBlue,
linkcolor=MyDarkBlue,
urlcolor=MyDarkBlue,
pdfauthor={Christian S\"amann, Martin Wolf},
pdftitle={Supersymmetric Yang-Mills Theory as Higher Chern-Simons Theory},
pdfsubject={hep-th},
breaklinks=true
}

\let\fn\footnote
\renewcommand{\footnote}[1]{\linespread{1.1}\fn{#1}\linespread{1.29}}

\makeatletter
\newcommand{\xRightarrow}[2][]{\ext@arrow 0359\Rightarrowfill@{#1}{#2}}
\renewcommand{\section}{\@startsection
{section}{1}{\z@}{-3.5ex plus -1ex minus
    -.2ex}{2.3ex plus .2ex}{\bf\mathversion{bold} }}
\renewcommand{\subsection}{\@startsection{subsection}{2}{\z@}{-3.25ex
plus -1ex minus
   -.2ex}{1.5ex plus .2ex}{\bf\mathversion{bold} }}
\renewcommand{\subsubsection}{\@startsection{subsubsection}{3}{-2.45ex}{-3.25ex
\makeatother
plus -1ex minus -.2ex}{1.5ex plus .2ex}{\it }}
\renewcommand{\thesection}{\arabic{section}}
\renewcommand{\thesubsection}{\arabic{section}.\arabic{subsection}}
\renewcommand{\@seccntformat}[1]{\@nameuse{the#1}.~~}

\renewcommand{\theequation}{\thesection.\arabic{equation}}
\makeatletter \@addtoreset{equation}{section}

\renewcommand*\l@section{\@dottedtocline{1}{0em}{2em}}
\renewcommand*\l@subsection{\@dottedtocline{2}{2em}{2.4em}}
\renewcommand*\l@subsubsection{\@dottedtocline{4}{3.8em}{3.7em}}

\renewcommand\tableofcontents{%
    \section*{\large\contentsname
        \@mkboth{%
          \MakeUppercase\contentsname}{\MakeUppercase\contentsname}}%
       {\baselineskip=15pt plus 2pt minus 1pt
    \@starttoc{toc}}%
}

\newcommand{\acknowledgements}{\section*{Acknowledgements}
\addcontentsline{toc}{section}{Acknowledgements}}

\newcommand{\datamanagement}{\section*{Data Management}
\addcontentsline{toc}{section}{Data Management}}

\newcommand{\appendices}{
\section*{Appendix}\label{appendices}\setcounter{subsection}{0}
\addcontentsline{toc}{section}{Appendix}
\setcounter{equation}{0}
\setcounter{thm}{0}
\makeatletter
\renewcommand{\theequation}{\Alph{subsection}.\arabic{equation}}
\renewcommand{\thesubsection}{\Alph{subsection}}
\renewcommand{\thethm}{\Alph{subsection}.\arabic{thm}}
\@addtoreset{equation}{subsection}
\@addtoreset{thm}{subsection}
\makeatother
}

\newtheorem{thm}{Theorem}[section]
\renewcommand{\thethm}{\thesection.\arabic{thm}}

\newtheorem{theorem}[thm]{Theorem}

\def\periodb#1{\setbox0=\hbox{$#1$}#1\hskip-\wd0\hbox to\wd0{-}}

\newcommand{\unit}{\mathbbm{1}}   			

\newcommand{\CC}{\mathcal{C}}

\newcommand{\CF}{\mathcal{F}}

\newcommand{\CG}{\mathcal{G}}
\newcommand{\CCG}{\mathscr{G}}
\newcommand{\CH}{\mathcal{H}}

\newcommand{\CN}{\mathcal{N}}

\newcommand{\CO}{\mathcal{O}}

\newcommand{\frg}{\mathfrak{g}}				
\newcommand{\frh}{\mathfrak{h}}				

\newcommand{\FR}{\mathbbm{R}}     			
\newcommand{\FC}{\mathbbm{C}}     			

\newcommand{\NN}{\mathbbm{N}}     			
\newcommand{\RZ}{\mathbbm{Z}}     			
\newcommand{\PP}{{\mathbbm{P}}}    			

\newcommand{\dd}{\mathrm{d}}     			
\newcommand{\dpar}{\partial}     			
\newcommand{\dparb}{{\bar{\partial}}}     		
\newcommand{\eps}{{\varepsilon}}			

\newcommand{\ald}{{\dot{\alpha}}}     			
\newcommand{\bed}{{\dot{\beta}}}

\newcommand{\eand}{{~~~\mbox{and}~~~}}     		
\newcommand{\ewith}{{~~~\mbox{with}~~~}}
\newcommand{\efor}{{~~~\mbox{for}~~~}}

\newcommand{\der}[1]{\frac{\dpar}{\dpar #1}}   		
\newcommand{\tr}{\,\mathrm{tr}\,}     			

\newcommand{\au}{\mathfrak{u}}

\newcommand{\sAd}{\mathsf{Ad}}

\newcommand{\sLie}{\mathsf{Lie}}

\newcommand{\sG}{\mathsf{G}}

\newcommand{\sL}{\sfL}
\newcommand{\sA}{\mathsf{A}}

\newcommand{\remark}[1]{}     				

\def\tyng(#1){\hbox{\tiny$\yng(#1)$}}			
\def\tyoung(#1){\hbox{\tiny$\young(#1)$}}			

\newcommand{\sfL}{\mathsf{L}}

\begin{document}

\begin{titlepage}

\setcounter{page}{0}
\renewcommand{\thefootnote}{\fnsymbol{footnote}}

\begin{flushright}
 EMPG--17--02\\ DMUS--MP--17/02
\end{flushright}

\begin{center}

{\LARGE\textbf{\mathversion{bold}Supersymmetric Yang--Mills Theory\\[-2pt] as Higher Chern--Simons Theory}\par}

\vspace{1cm}

{\large
Christian S\"amann$^{a}$ and Martin Wolf$^{\,b}$
\footnote{{\it E-mail addresses:\/}
\href{mailto:c.saemann@hw.ac.uk}{\ttfamily c.saemann@hw.ac.uk}, 
\href{mailto:m.wolf@surrey.ac.uk}{\ttfamily m.wolf@surrey.ac.uk}
}}

\vspace{.3cm}

{\it
$^a$ Maxwell Institute for Mathematical Sciences\\
Department of Mathematics,
Heriot--Watt University\\
Edinburgh EH14 4AS, United Kingdom\\[.3cm]

$^b$
Department of Mathematics,
University of Surrey\\
Guildford GU2 7XH, United Kingdom
}

\vspace{2cm}

{\bf Abstract}
\end{center}
\vspace{-.5cm}
\begin{quote}
We observe that the string field theory actions for the topological sigma models describe higher or categorified Chern--Simons theories. These theories yield dynamical equations for connective structures on higher principal bundles. As a special case, we consider holomorphic higher Chern--Simons theory on the ambitwistor space of four-dimensional space-time. In particular, we propose a higher ambitwistor space action functional for maximally supersymmetric Yang--Mills theory. 

\vfill\noindent 26th July 2017

\end{quote}

\setcounter{footnote}{0}\renewcommand{\thefootnote}{\arabic{thefootnote}}

\end{titlepage}

\tableofcontents
\bigskip
\bigskip
\hrule
\bigskip
\bigskip

\section{Introduction and results}

In~\cite{Witten:9207094}, Witten used his proposal for open string field theory~\cite{Witten:1986qs} to link topological string theory to Chern--Simons theory. This formulation of string field theory, however, has been argued to be singular or incomplete~\cite{Wendt:1987zh}, see also~\cite{Erler:2013xta}. To remedy this problem, Erler et al.~\cite{Erler:2013xta} resolved the singularity of the cubic vertex in Witten's formulation resulting in the structure of an $A_\infty$-algebra on the Hilbert space of string fields. For earlier accounts on formulating open string field theory using (parts of) $A_\infty$-algebras, see also~\cite{Hata:1986jd,Hata:1986kj,Gaberdiel:1997ia,Zwiebach:1997fe,Nakatsu:2001da,Iimori:2013kha}. Similarly, closed string field theory has an underlying $L_\infty$-algebra structure~\cite{Zwiebach:1992ie}. The corresponding string field theory actions yield the homotopy Maurer--Cartan equations for both $A_\infty$-algebras and $L_\infty$-algebras.

Particularly interesting Hilbert spaces to feed into this string field theory formalism are those of the topological A model and B model~\cite{Witten:1991zz}, that is, the de Rham complex on a special Lagrangian submanifold of a Calabi--Yau manifold\footnote{We shall restrict ourselves to Calabi--Yau manifolds for simplicity.} $X$ as well as the Dolbeault complex on $X$, respectively. The first aim of this paper is to present the resulting actions and equations of motion.

For $X$ a three-dimensional Calabi--Yau manifold, Witten showed~\cite{Witten:9207094} that the topological A model and B model reduce to Chern--Simons theory and holomorphic Chern--Simons theory, respectively. We observe that this result is not affected by the generalisation to $A_\infty$-algebras. For higher-dimensional Calabi--Yau manifolds, however, we obtain higher or categorified versions of Chern--Simons theory and its holomorphic variant. Higher Chern--Simons theories have been studied in various settings before, see e.g.~\cite{Sati:0801.3480,Fiorenza:2010mh,Fiorenza:2011jr,Soncini:2014ara,Zucchini:2015ohw,Ritter:2015ymv}. They describe connective structures on categorified principal $n$-bundles which have Lie $n$-groups as their structure groups with $n=\dim_\FC(X)-2$. For instance in the case of a four-dimensional Calabi--Yau manifold, we are led to a principal 2-bundle with a structure Lie 2-group, whose connective structure contains besides a 1-form gauge potential also a 2-form gauge potential.

A proposal of generalising Chern--Simons theory to higher form theories as a description of the A model has been made before by Schwarz\cite{Schwarz:2005rv} but the underlying assignment of Chan--Paton factors was rather ad hoc and limited to ordinary Lie algebras. This action was subsequently used to construct a Batalin--Vilkovisky action functional on harmonic superspace for maximally supersymmetric Yang--Mills (MSYM) theory in ten and lower dimensions \cite{Movshev:2003ib,Movshev:2004aw}. A first aim of this paper is to show how to incorporate the full $A_\infty$-algebra structure and how this arises in a natural way from string field theory. 

The topological A model and B model are usually studied for three-dimensional target spaces since it is known that correlation functions in topological string theory for other dimensions are mostly trivial. There is, however, one particularly interesting higher-dimensional target space, namely the ambitwistor space relevant in the twistor string theory description of MSYM theory in four dimensions.

Twistor string theory~\cite{Witten:2003nn} has led to new ways of computing Yang--Mills scattering amplitudes and in some procedures, an explicit action functional is desirable. On the usual extension of Penrose's twistor space to a Calabi--Yau supermanifold $P^{3|4}$, this is not an issue as the three-dimensional bosonic part of $P^{3|4}$ allows for using the ordinary holomorphic Chern--Simons action functional ~\cite{Witten:2003nn,Mason:2005zm,Boels:2006ir}. In this context, the holomorphic Chern--Simons action yields an action functional for self-dual MSYM theory which can be complemented to a twistor space action for full MSYM theory ~\cite{Witten:2003nn,Mason:2005zm,Boels:2006ir}. On the ambitwistor space with its five-dimensional bosonic part, however, the holomorphic Chern--Simons action needs to be replaced, and up to now, a holomorphic BF-type action or the more involved constructions of~\cite{Mason:2005kn} have been used. 

In the second part of this paper, we propose an alternative. The action functional of higher holomorphic Chern--Simons theory offers itself very naturally on ambitwistor space to be used for twistor string theory. We argue that even the additional degrees of freedom present in the higher form fields are gauge trivial on-shell. Therefore we obtain, at least at the classical level, an equivalence between higher holomorphic Chern--Simons theory on the ambitwistor space and MSYM theory in four dimensions.

There are a number of future directions arising from our results. First of all, it would be interesting to examine the use of higher holomorphic Chern--Simons theory within twistor string theory and the computation of MSYM scattering amplitudes. Secondly, it would be interesting to study an extension of our results for the topological A model and B model to topological string theory by coupling them to topological gravity. One should directly perform this analysis on supermanifolds, using e.g.\ the techniques of~\cite{Jia:2016jlo,Jia:2016rdn}. Particularly interesting here are the cancellation of the various anomalies and the existence and meaning of D-branes. These should then be linked to our generalised notion of Chan--Paton factors. 

\section{String field theory and homotopy Maurer--Cartan equations}

\subsection{\texorpdfstring{$A_\infty$}{A_infinity}-algebras and \texorpdfstring{$L_\infty$}{L_infinity}-algebras}

Roughly speaking, an $A_\infty$-algebra generalises the matrix product in a matrix algebra to a non-associative structure while an $L_\infty$-algebra generalises the commutator in a matrix algebra to a structure violating the Jacobi identity. Just as the commutator arises from the anti-symmetrisation of the matrix product, the anti-symmetrisation of the products in an $A_\infty$-algebra yields an $L_\infty$-algebra. The modern and elegant approach to the definition of such algebras uses a dual formulation in terms of differential graded algebras. This, however, would lead us to far afield from the main thread of development and we merely refer to~\cite{Stasheff:1992bb} for details, see also~\cite{Stasheff:1963aa,Lada:1992wc} for more detailed accounts of strong homotopy algebras. We shall adopt the conventions of \cite{Penkava:1994mu}.

An {\it $A_\infty$-algebra} or {\it strong homotopy associative algebra} is a $\RZ$-graded vector space $\sA=\oplus_{p\in \RZ} \sA_p$ together with graded, multilinear maps
\begin{equation}
 m_i\,:\, \otimes^i\sA\ \rightarrow\ \sA
\end{equation}
of degree~$2-i$ for $i\in\NN$ subject to the {\it higher homotopy relations}
\begin{equation}\label{eq:homotopy_relations}
 \sum_{\substack{r+s+t=i\\ r,t\in\NN_0\\ s\in\NN}} (-1)^{rs+t}\, m_{r+1+t}\circ (\underbrace{\unit\otimes\cdots\otimes\unit}_{\scriptscriptstyle r-{\rm times}} \otimes\, m_s\otimes\underbrace{\unit\otimes\cdots\otimes\unit}_{\scriptscriptstyle t-{\rm times}})\ =\ 0
\end{equation}
for $i\in\NN$. Denoting the degree of an element $a\in \sA$ by $|a|$, we have\footnote{Clearly, these equations only make sense on elements $a_i\in\sA$ with homogeneous degree. Otherwise, one simply needs to expand the equations linearly. We shall always suppress this in the following.} $|m_i(a_1,\ldots,a_i)|=2-i+|a_1|+\cdots+|a_i|$ and the following explicit form of these relations for $i\leq 3$:
\begin{equation}\label{eq:homotopy_relations_Ainfty}
\begin{aligned}
 i=1:~~&m_1(m_1(a_1))\ =\ 0~,\\
 i=2:~~&m_1(m_2(a_1,a_2))-m_2(m_1(a_1),a_2)-(-1)^{|a_1|}m_2(a_1,m_1(a_2))\ =\ 0~,\\
 i=3:~~&m_1(m_3(a_1,a_2,a_3))-m_2(m_2(a_1,a_2),a_3)+m_2(a_1,m_2(a_2,a_3))\,+\\
 &\hspace{.5cm}+m_3(m_1(a_1),a_2,a_3)+(-1)^{|a_1|}m_3(a_1,m_1(a_2),a_3)\,+\\
 &\hspace{1cm}+(-1)^{|a_1|+|a_2|}m_3(a_1,a_2,m_1(a_3))\ =\ 0
\end{aligned}
\end{equation}
for $a_1,a_1,a_3\in\sA$, where signs arise from permuting odd elements of $\sA$ past products with odd degrees. The first relation evidently states that $m_1$ is nil-quadratic and the second relation implies its compatibility with the product $m_2$ in terms of a Leibniz rule, and so $m_1$ is a differential. The third relation determines the controlled violation of associativity of $m_2$. 

The $A_\infty$-algebras of interest are usually nontrivial only in degrees~$p\geq 0$ or $p \leq 0$, and we shall encounter both situations. If an $A_\infty$-algebra $\sA$ is trivial except for degrees~$p$ with $-(n-1)\leq p\leq 0$ or $0\leq p\leq n-1$, we call it an {\it $n$-term $A_\infty$-algebra}. 

When formulating action principles, we shall also need an appropriate notion of an inner product on an $A_\infty$-algebra. The first formulation of inner products on $A_\infty$-algebras we are aware of was given in~\cite{Kontsevich:1992aa}, see also~\cite{Kajiura:0306332} and in particular~\cite{0821843621} for a definition from the cyclic operad. An {\it inner product} on an $A_\infty$-algebra $\sA$ over $\FR$ is an even, graded symmetric, non-degenerate, bilinear map
\begin{equation}
 \langle-,-\rangle_\sA\,:\,\sA\odot \sA\ \rightarrow\ \FR~,
\end{equation}
which is cyclic in the sense that
\begin{equation}
 \langle a_1,m_i(a_2,\ldots,a_{i+1})\rangle_\sA\ =\ (-1)^{i+i(|a_1|+|a_{i+1}|)+|a_{i+1}|\sum_{j=1}^{i}|a_j|}\langle a_{i+1},m_i(a_1,a_2,\ldots,a_{i})\rangle_\sA
\end{equation}
for $i\in\NN$ and $a_1,\ldots,a_{i+1}\in \sA$. An inner product $A_\infty$-algebra is also called a {\it cyclic $A_\infty$-algebra}. Note that on an $n$-term $A_\infty$-algebra $\sA=\sA_{-n+1}\oplus \cdots \oplus \sA_0$, this bilinear map induces an isomorphism $\sA_{p}\rightarrow \sA^*_{-n+1-p}$ for $-n+1\leq p\leq 0$, because it is non-degenerate.

The definition of $L_\infty$-algebras is very similar. An {\it $L_\infty$-algebra} or {\it strong homotopy Lie algebra} $\sL$ is a $\RZ$-graded vector space $\sL=\oplus_{p\in\RZ}\sL_p$ endowed with graded, totally anti-symmetric, linear maps
\begin{equation}
 \mu_i\,:\, \wedge^i\sL\ \rightarrow\ \sL
 \end{equation}
for $i\in\NN$ of degree~$2-i$ subject to the {\it higher homotopy Jacobi identities}
\begin{equation}\label{eq:homotopyJacobi}
 \sum_{r+s=i}\sum_{\sigma }\chi(\sigma;\ell_1,\ldots,\ell_{r+s})(-1)^{s}\mu_{s+1}(\mu_r(\ell_{\sigma(1)},\ldots,\ell_{\sigma(r)}),\ell_{\sigma(r+1)},\ldots,\ell_{\sigma(r+s)})\ =\ 0
\end{equation}
for $i\in\NN$ and $\ell_1,\ldots,\ell_{r+s}\in \sL$. Here, the sum over $\sigma$ is taken over all $(r,s)$ {\it unshuffles} which consist of permutations $\sigma$ of $\{1,\ldots,r+s\}$ such that the first $r$ and the last $s$ images of $\sigma$ are ordered: $\sigma(1)<\cdots<\sigma(r)$ and $\sigma(r+1)<\cdots<\sigma(r+s)$. Moreover, $\chi(\sigma;\ell_1,\ldots,\ell_i)$ is the {\it graded Koszul sign} defined by
\begin{equation}
 \ell_1\wedge \ldots \wedge \ell_i\ =\ \chi(\sigma;\ell_1,\ldots,\ell_i)\,\ell_{\sigma(1)}\wedge \ldots \wedge \ell_{\sigma(i)}
\end{equation}
in the free graded algebra. The higher homotopy Jacobi identities~\eqref{eq:homotopyJacobi} for $i=1$ and $i=2$ state that $\mu_1$ is a differential which is compatible with the product $\mu_2$ just as in the case of $A_\infty$-algebras. The corresponding relation for $i=3$ describes the controlled violation of the graded Jacobi identity.

We again consider {\it $n$-term $L_\infty$-algebras} which have underlying graded vector space concentrated in degrees~$p$ with $-(n-1)\leq p\leq 0$ or $0\leq p\leq n-1$. At least $L_\infty$-algebras of the form $\sL=\sL_{-1}\oplus\sL_0$ are categorically equivalent to semistrict Lie $2$-algebras~\cite{Baez:2003aa}. In the following, we shall often simply speak of Lie $n$-algebras when we mean an $L_\infty$-algebra of the form $\sL=\sL_{-n+1}\oplus \cdots \oplus \sL_{-1}\oplus \sL_{0}$.

Finally, we note that there is a functor from the category of $A_\infty$-algebras to that of $L_\infty$-algebras~\cite{Lada:1994mn} which on objects reads as 
\begin{equation}\label{eq:functor_A_to_L}
 \mu_i(\ell_1,\ldots,\ell_i)\ =\ \sum_{\sigma}\chi(\sigma;\ell_1,\ldots,\ell_i)m_i(\ell_{\sigma(1)},\ldots,\ell_{\sigma(i)})
\end{equation}
with the sum running over all permutations. In other words, the total graded anti-symmetrisation of the products in an $A_\infty$-algebra  yields an $L_\infty$-algebra with the same underlying graded vector space. For this functor to extend to cyclic $A_\infty$-algebras, we define an {\it inner product} on an $L_\infty$-algebra $\sL$ over $\FR$ as an even, graded symmetric, non-degenerate, bilinear map
\begin{equation}
 \langle -,-\rangle_\sL\,:\,\sL\odot \sL\ \rightarrow\ \FR~,
\end{equation}
which is cyclic in the sense of
\begin{equation}
 \langle\ell_1,\mu_i(\ell_2,\ldots,\ell_{i+1})\rangle_\sL\ =\ (-1)^{i+i(|\ell_1|+|\ell_{i+1}|)+|\ell_{i+1}|\sum_{j=1}^{i}|\ell_j|}\langle\ell_{i+1},\mu_i(\ell_1,\ldots,\ell_{i})\rangle_\sL
\end{equation}
for $i\in\NN$.

\subsection{Maurer--Cartan equations and action principles}\label{ssec:MCEqn}

The Maurer--Cartan equation $\dd \omega+\omega\wedge \omega=0$ for a matrix Lie algebra valued differential one-form $\omega$ naturally generalises to $A_\infty$-algebras $\sA=\oplus_{p\in\RZ}\sA_p$. In particular, {\it Maurer--Cartan elements} of $\sA$ are elements $a\in \sA_1$ satisfying\footnote{Elements of degree different from 1 in $\sA$ should be regarded as ghosts, cf.\ the discussion in~\cite{Witten:9207094} or, e.g.,~\cite{Zucchini:2017nax} for the case of higher gauge theory.} the {\it homotopy Maurer--Cartan equation},
\begin{equation}\label{eq:hMC}
 \sum_{i\in\NN} m_i(a,\ldots,a)\ =\ m_1(a)+m_2(a,a)+\cdots \ = \ 0~.
\end{equation}
We will now look at homotopies between Maurer--Cartan elements, which will lead us to a higher form of gauge symmetry.

It is not very difficult to see that the tensor product of a differential graded vector space with an $A_\infty$-algebra or an $L_\infty$-algebra carries again a natural $A_\infty$-algebra or $L_\infty$-algebra structure, cf.\ e.g.~\cite{Jurco:2014mva}. Let $\sA$ be an $A_\infty$-algebra with products $m_i$, and $V=\oplus_{p\in\RZ} V_p$ be a differential graded vector space with differential $\dd$. We then define an $A_\infty$-algebra 
\begin{equation}
 \hat \sA\ :=\ V\otimes \sA \ :=\ \bigoplus_{p\in\RZ} \hat \sA_p\ewith \hat \sA_p\ :=\ \bigoplus_{i+j=p}~V_i\otimes \sA_j
\end{equation}
such that $|v\otimes a|=|v|+|a|$ for $v\in V$ and $a\in \sA$. The higher products $\hat m_i$ on $\hat \sA$ are defined as 
\begin{equation}
\begin{aligned}
 \hat m_1(v_1 \otimes a_1)\ &:=\ \dd v_1 \otimes a_1+(-1)^{|v_1|}v_1\otimes m_1(a_1)~,\\[5pt]
 \hat m_i(v_1\otimes a_1,\ldots,v_i\otimes a_i)\ &:=\ (-1)^{i\sum_{j=1}^i|v_j|+\sum_{j=0}^{i-2}|v_{i-j}|\sum_{k=1}^{i-j-1}|a_k|}\,\times\\[-5pt]
 &\kern2cm\times(v_1\wedge \ldots \wedge v_i)\otimes m_i(a_1,\ldots,a_i)
\end{aligned}
\end{equation}
for $i\geq 2$ and $v_1,\ldots,v_i\in V$ and $a_1,\ldots,a_i\in \sA$. One readily verifies that these products satisfy the relation~\eqref{eq:homotopy_relations}. We also note that in the special case of $V$ being the de Rham complex $(\Omega^\bullet(M),\dd)$ on a manifold $M$,
\begin{equation}\label{eq:hat_inner_product}
 \langle \omega_1\otimes a_1,\omega_2\otimes a_2\rangle_{\hat \sA}\ :=\ (-1)^{|\omega_2| |a_1|}\int_M \omega_1\wedge \omega_2~\langle a_1,a_2\rangle_\sA
\end{equation}
 defines an inner product on $\hat \sA:=\Omega^\bullet(M,\sA):=\Omega^\bullet(M)\otimes\sA$ for $\omega_{1,2}\in \Omega^\bullet(M)$ and $a_{1,2}\in \sA$ where $\langle-,-\rangle_\sA$ is a fixed inner product on $\sA$.

Homotopies between Maurer--Cartan elements in $\sA$ are now described by Maurer--Cartan elements in $\hat \sA=\Omega^\bullet(I,\sA)$, where $I=[0,1]\subset \FR$. An element of degree~1 in $\hat \sA$ is written as
\begin{equation}
 \hat a\ =\ a+\dd t~\lambda~,
\end{equation}
where $t\in I$, $a\in \Omega^0(I,\sA_1)$ and $\lambda\in \Omega^0(I,\sA_0)$. The homotopy Maurer--Cartan equation now determines $\left.\der{t} a(t)\right|_{t=0}$ which can be identified with an infinitesimal higher gauge transformation $\delta a$ parametrised by $\lambda=\lambda(0)\in \sA_0$,
\begin{equation}\label{eq:gauge_trafos_A_infty}
\begin{aligned}
 \delta a\ &:=\ \sum_{i\in\NN} \big[(-1)^{i-1} m_i(\lambda,\underbrace{a,\ldots,a}_{\scriptscriptstyle (i-1)-{\rm times}})\,+\\
 &\kern2cm + (-1)^{i-2}m_i(a,\lambda,\underbrace{a,\ldots,a}_{\scriptscriptstyle (i-2)-{\rm times}})+\cdots+m_i(\underbrace{a,\ldots,a}_{\scriptscriptstyle (i-1)-{\rm times}},\lambda) )\big]~.
 \end{aligned}
\end{equation}

An action functional for equation \eqref{eq:hMC} which is invariant under~\eqref{eq:gauge_trafos_A_infty} is
\begin{equation}\label{eq:action_A_infty}
 S\ :=\ \sum_{i\in\NN} \frac{1}{i+1} \langle a,m_i(a,\ldots,a)\rangle_\sA~,
\end{equation}
where $a\in \sA_1$. Note that the action~\eqref{eq:action_A_infty} is precisely the action found in~\cite{Erler:2013xta} on the $A_\infty$-algebra formed by the Hilbert space $\CH$ of string states. The higher products $m_i$ which render the Hilbert space $\CH$ into an $A_\infty$-algebra are then computed from the various $i$-point functions.

In closed string field theory, the natural analogue for $L_\infty$-algebras is used~\cite{Zwiebach:1992ie}, see also~\cite{Lada:1992wc}. Here, the action, equation of motion and gauge transformations are
\begin{equation}
\begin{gathered}
 S\ :=\ \sum_{i\in\NN} \frac{1}{(i+1)!}\langle\ell,\mu_i(\ell,\ldots,\ell)\rangle_\sL~,~~~\sum_{i\in\NN}\frac{1}{i!}\mu_i(\ell,\ldots,\ell)\ =\ 0~,\\
 \delta \ell\ :=\ \sum_{i\in\NN} \frac{1}{(i-1)!}\mu_i(\ell,\ldots,\ell,\lambda)
\end{gathered}
\end{equation}
for $\ell\in\sL_1$ and $\lambda\in\sL_0$ of an $L_\infty$-algebra $\sL$. Clearly, these equations arise from the functor taking $A_\infty$-algebras to $L_\infty$-algebras.

Finite gauge transformations can be obtained by integrating the infinitesimal ones using the techniques of~\cite{Henriques:2006aa}. The integration of the $L_\infty$-algebra $\sL$ leads to a higher or categorified Lie group $\CCG$. Particularly interesting is the case in which the elements $\ell\in \sL$ form a connective structure on a categorified principal $\CCG$-bundle. The theory of such bundles has been developed to some detail in the literature, see e.g.~\cite{Fiorenza:2010mh,Nikolaus:1207ab} as well as~\cite{Jurco:2016qwv} for an easily accessible description. Note that action~\eqref{eq:action_A_infty} captures only the cases of trivial principal $\CCG$-bundles, the general discussion is found, e.g., in~\cite{Sati:0801.3480}.

\section{From topological sigma models to higher Chern--Simons theory}

Let us now apply the previous formalism to the special case of topological sigma models. The detailed construction of these models is found in~\cite{Witten:1991zz,Witten:9207094}, see also~\cite{Witten:1988xj}. 

\subsection{A model and B model}

The two topological sigma models we are interested in are called the {\it A model} and the {\it B model} and they describe maps from a Riemann surface $\Sigma$ to a target space $X$. Although this is stronger than necessary, we shall assume that $X$ is a Calabi--Yau manifold, by which we mean\footnote{For ordinary manifolds, this is equivalent to the first Chern class vanishing, the existence of a Ricci flat K\"ahler metric etc. In the case of supermanifolds, however, this condition appears to be weaker, see e.g.~\cite{Rocek:2004bi}.} a (potentially non-compact) K\"ahler (super)manifold with a globally defined holomorphic measure. 

Besides the bosonic fields encoded in the map $\phi:\Sigma\rightarrow X$, we also have fermions which are sections of ordinary vector bundles due to the topological twist. In the case of the A model we have fermions $\chi$ which are sections of $\phi^*TX$ as well as fermions $\psi$ which are sections of $\Omega^1(\Sigma,\phi^*TX)$ satisfying an additional constraint. In the case of the B model, we have fermions $\eta$ and $\theta$ which are sections of $\phi^*T^{0,1}X$ as well as a field $\rho$ which is a section of $\Omega^1(\Sigma, \phi^*T^{1,0}X)$, where $TX\otimes\FC\cong T^{1,0}X\oplus T^{0,1}X$ is the decomposition into the holomorphic and anti-holomorphic tangent bundle.

For both models, a topological Lagrangian is readily written down, which can be quantised. In the case of the A model, let $M$ be a special Lagrangian submanifold in $X$ which contains boundaries of the image of the worldsheet $\Sigma$ under $\phi$. The BRST cohomology of the A model can then be represented as functions of $\phi$ and $\chi$ restricted to $M$. Since $\chi$ is fermionic, these functions are identified with differential forms on $M$. Moreover, the BRST operator is mapped to the de Rham differential $\dd$ on $M$. Altogether, the Hilbert space of the topological A model with boundary $M$ is described by the de Rham complex 
\begin{equation}
 \CC^\infty(M)\ \xrightarrow{~\dd~}\ \Omega^{1}(M)\ \xrightarrow{~\dd~}\ \Omega^{2}(M)\ \xrightarrow{~\dd~}\ \cdots
\end{equation}
on $M$. For the B model, a similar line of arguments shows that the BRST cohomology of the B model is identified with the Dolbeault complex 
\begin{equation}
 \CC^\infty(X)\ \xrightarrow{~\dparb~}\ \Omega^{0,1}(X)\ \xrightarrow{~\dparb~}\ \Omega^{0,2}(X)\ \xrightarrow{~\dparb~}\ \cdots~,
\end{equation}
where $\dparb$ denotes the anti-holomorphic exterior derivative. Here, the boundaries of the worldsheet $\Sigma$ are restricted to lie in complex submanifolds of $X$.

If we want to regard the boundary of $\Sigma$ as being contained in a single D-brane, then this is the complete story. To capture stacks of multiple D-branes, however, we need to introduce Chan--Paton factors on the boundaries. This amounts to tensoring the de Rham complex by the endomorphism bundle of a flat vector bundle or the Dolbeault complex by the endomorphism bundle of a holomorphic vector bundle, respectively. These bundles, in turn, are described by Chern--Simons theory and holomorphic Chern--Simons theory, respectively~\cite{Witten:9207094}. In this context, we shall only be interested in space-filling D-branes.

\subsection{Ordinary Chan--Paton factors and higher Chern--Simons theory}\label{ssec:ordinary_Chan_Paton}

The Hilbert spaces of the topological sigma models with target space $X$ form now the underlying graded vector spaces of two $A_\infty$-algebras. Let $d:=\dim_\FR(M)=\dim_\FC(X)$ and $n=d-2$. The ordinary assignment of Chan--Paton factors leads to the cyclic $n$-term $A_\infty$-algebra $\sA=\sA_{-n+1}\oplus \cdots \oplus \sA_0$ with $\sA_i\cong \frg$ for some matrix Lie algebra $\frg$. The only non-trivial product is $m_2$, induced as follows from the matrix product\footnote{Further restrictions can also be used, e.g.\ $m_2(a_1,a_2)=0$ unless $|a_1|=0$ or $|a_2|=0$.} in $\frg$:
\begin{equation}
m_2(a_1,a_2)\ :=\ \begin{cases} a_1a_2\in \sA_{|a_1|+|a_2|} & \efor |a_1|+|a_2|>-n~,\\
0 &\quad\mbox{else}~.
\end{cases}
\end{equation}
The cyclic inner product is induced by the Hilbert--Schmidt inner product through
\begin{equation}
\langle a_1,a_2\rangle_\sA\ :=\ \begin{cases} \tr(a_1^\dagger a_2) &\efor |a_1|+|a_2|\ =\ -n+1~,\\
0 &\quad\mbox{else}~.
\end{cases}
\end{equation}

Tensoring the de Rham and Dolbeault complexes by $\sA$ as explained in Section~\ref{ssec:MCEqn} leads to the two $A_\infty$-algebras for the A model and B model,
\begin{equation}\label{eq:BRST_cohomologies}
\begin{gathered}
    \Omega^\bullet(M,\sA)\ :=\ \bigoplus_{p\in\NN_0}\Omega^\bullet_p(M,\sA)\ewith \Omega^\bullet_p(M,\sA)\ :=\ \bigoplus_{\substack{i+j=p\\ 0\leq i\leq d\\-n+1\leq j\leq 0}} \Omega^{i}(M,\sA_{j})~,\\
      \Omega^{0,\bullet}(X,\sA)\ :=\ \bigoplus_{p\in\NN_0}\Omega^{0,\bullet}_p(X,\sA)\ewith \Omega^{0,\bullet}_p(X,\sA)\ :=\ \bigoplus_{\substack{i+j=p\\ 0\leq i\leq d\\-n+1\leq j\leq 0}} \Omega^{0,i}(X,\sA_{j})~,
     \end{gathered}
\end{equation}
respectively. The non-trivial higher products on $\Omega^\bullet(M,\sA)$ are linear extension of 
\begin{subequations}
\begin{equation}
\begin{aligned}
\hat m_1(\omega_1\otimes a_1)\ &:=\ \dd\omega_1\otimes a_1~,\\
\hat m_2(\omega_1\otimes a_1,\omega_2\otimes a_2)\ &:=\ (-1)^{|a_1||\omega_2|}(\omega_1\wedge \omega_2)\otimes a_1a_2~,
\end{aligned}
\end{equation}
where $\omega_{1,2}\in \Omega^\bullet(M)$ and $a_{1,2}\in\sA$ while for $\Omega^{0,\bullet}(X,\sA)$, we have
\begin{equation}
\begin{aligned}
\hat m_1(\omega_1\otimes a_1)\ &:=\ \dparb \omega_1\otimes a_1~,\\
\hat m_2(\omega_1\otimes a_1,\omega_2\otimes a_2)\ &:=\ (-1)^{|a_1||\omega_2|}(\omega_1\wedge \omega_2)\otimes a_1a_2~,
\end{aligned}
\end{equation}
\end{subequations}
where $\omega_{1,2}\in \Omega^{0,\bullet}(X)$ and $a_{1,2}\in\sA$. Furthermore, the cyclic structure~\eqref{eq:hat_inner_product} is 
\begin{subequations}\label{eq:cyclic_naive}
\begin{equation} 
\langle\omega_1\otimes a_1,\omega_2\otimes a_2\rangle_{\hat \sA}\ :=\ (-1)^{|a_1||\omega_2|}\int_M \omega_1\wedge \omega_2\,\tr(a_1^\dagger a_2)
\end{equation}
on $\Omega^\bullet(M,\sA)$, while on $\Omega^{0,\bullet}(M,\sA)$, we choose
\begin{equation}
 \langle\omega_1\otimes a_1,\omega_2\otimes a_2\rangle_{\hat \sA}\ :=\ (-1)^{|a_1||\omega_2|} \int_X\Omega^{d,0}\wedge \omega_1\wedge \omega_2\,\tr(a_1^\dagger a_2)~,
\end{equation}
\end{subequations}
where $\Omega^{d,0}$ is the holomorphic volume form on $X$. In order to simplify our discussion, we shall restrict ourselves to $\Omega^\bullet(M,\sA)$ in the remainder of this section. The discussion for $\Omega^{0,\bullet}(X,\sA)$ is analogous, and we shall return to it in Section~\ref{sec:YMhCS}.

We can now plug the $A_\infty$-algebra $\Omega^\bullet(M,\sA)$ into the homotopy Maurer--Cartan action~\eqref{eq:action_A_infty} for a degree~1 element $\hat a=\hat a_1+\hat a_2+\cdots$ with $\hat a_i$ an $\sA_{-i+1}$-valued differential $i$-form. The result is 
\begin{equation}\label{eq:simpl_hCS}
\begin{aligned}
S\ &=\ \int_M \tr\left\{ \frac12 \hat a^\dagger \wedge \dd \hat a+ \frac{1}{3}\hat a^\dagger\wedge \hat a\wedge \hat a\right\}\\
  &=\ \int_M \tr\left\{ \frac12\sum_{i+j=d-1} \hat a_{i}^\dagger\wedge \dd \hat a_{j}+\frac{1}{3}\sum_{i+j+k=d} \hat a_{i}^\dagger\wedge \hat a_{j}\wedge \hat a_{k}\right\}~.
 \end{aligned}
\end{equation}
Note that for $d=3$, this action reduces to ordinary Chern--Simons theory, as expected from~\cite{Witten:9207094}. For $d>3$, the action~\eqref{eq:simpl_hCS} was suggested by Schwarz~\cite{Schwarz:2005rv} as an action for the A model with target spaces of dimension larger than three.

\subsection{Higher Chan--Paton factors}\label{ssec:HigherChanPaton}

Let us now generalise the notion of Chan--Paton factors to proper $A_\infty$-algebras as suggested by string field theory. That is, we start from a full $n$-term $A_\infty$-algebra $\sA=\sA_{-n+1}\oplus \cdots \oplus\sA_0$ with higher products $m_i$ and tensor it by the de Rham complex as explained in Section~\ref{ssec:MCEqn}. We obtain 
\begin{equation}
 \begin{aligned}
  \Omega^\bullet(M,\sA)\ :=\ \bigoplus_{p\in\NN_0}\Omega^\bullet_p(M,\sA)\ewith \Omega^\bullet_p(M,\sA)\ :=\ \bigoplus_{\substack{i+j=p\\ 0\leq i\leq d\\-n+1\leq j\leq 0}} \Omega^{i}(M,\sA_{j})
 \end{aligned}
\end{equation}
and higher products $\hat m_i$, which are linear extensions of the following ones:
\begin{equation}
\begin{aligned}
 \hat m_1(\omega_1 \otimes a_1)\ &:=\ \dd \omega_1 \otimes a_1+(-1)^{|\omega_1|}\omega_1\otimes m_1(a_1)~,\\[5pt]
 \hat m_i(\omega_1\otimes a_1,\ldots,\omega_i\otimes a_i)\ &:=\ (-1)^{i\sum_{j=1}^i|\omega_j|+\sum_{j=0}^{i-2}|\omega_{i-j}|\sum_{k=1}^{i-j-1}|a_k|}\,\times\\[-5pt]
 &\kern2cm\times(\omega_1\wedge \ldots \wedge \omega_i)\otimes m_i(a_1,\ldots,a_i)
\end{aligned}
\end{equation}
for $i\geq 2$ and $\omega_1,\ldots,\omega_i\in \Omega^\bullet(M)$ and $a_1,\ldots,a_i\in \sA$. The cyclic structure or inner product on $\Omega^\bullet(M,\sA)$ is given in~\eqref{eq:hat_inner_product}.

Let again $d=\dim_\FC(X)=\dim_\FR(M)$ and $\sA=\sA_{-d+3}\oplus\cdots\oplus\sA_0$ be a $(d-2)$-term $A_\infty$-algebra. The degree~1 element $\hat a\in\Omega^\bullet(M,\sA)$ appearing in the homotopy Maurer--Cartan action~\eqref{eq:action_A_infty} decomposes according to $\hat a=\hat a_{1}+\hat a_{2}+\cdots$, where $\hat a_{i}$ is an $A_{-i+1}$-valued differential $i$-form. Gauge transformations~\eqref{eq:gauge_trafos_A_infty} are encoded in elements of $\Omega^\bullet(M,\sA)$ of degree~0, and the higher curvatures of $\hat a$ that vanishes according to the equation of motion is an element of $\Omega^\bullet(M,\sA)$ of degree~2. 

\paragraph{\mathversion{bold}Case $d=3$.}
When $d=3$, a 1-term $A_\infty$-algebra such as $\sA:=\au(N)$ with $m_2$ the matrix product will do. A degree~1 element in $\Omega^\bullet(M,\sA)$ is a $\au(N)$-valued differential one-form $A$. Consequently, the homotopy Maurer--Cartan action~\eqref{eq:action_A_infty} 
becomes
\begin{equation}
 S\ =\ \int_M\tr\left\{\tfrac12 A\wedge \dd A+\tfrac13 A\wedge A\wedge A\right\}.
\end{equation}
We thus recover ordinary Chern--Simons theory as in~\cite{Witten:9207094}.

\paragraph{\mathversion{bold}Case $d=5$.}
Here we should consider a cyclic 3-term $A_\infty$-algebra $\sA$. The cyclicity of $\sA$ implies that 
\begin{equation}
 \sA\ =\ \sA_{-2}\oplus \sA_{-1}\oplus \sA_0\ :=\ \frg^* \oplus \frh\oplus \frg~,
\end{equation}
where the cyclic structure is simply given by the natural pairing, together with an inner product $\langle-,-\rangle_\frh$ on $\frh$,
\begin{equation}
 \langle a_{-2}+a_{-1}+a_0,b_{-2}+b_{-1}+b_0\rangle_\sA\ =\ b_{-2}(a_0)+\langle a_{-1},b_{-1}\rangle_{\frh}+a_{-2}(b_0)~,
\end{equation}
where $a_i,b_i\in \sA_i$. Consequently, $\Omega^\bullet(M,\sA)=\bigoplus_{p=0}^5\Omega^\bullet_p(M,\sA)$ is given by
\begin{equation}
 \begin{aligned}
  \Omega^\bullet_0(M,\sA)\ &=\ \Omega^0(M,\frg)\oplus\Omega^1(M,\frh)\oplus\Omega^2(M,\frg^*)~,\\
  \Omega^\bullet_1(M,\sA)\ &=\ \Omega^1(M,\frg)\oplus\Omega^2(M,\frh)\oplus\Omega^3(M,\frg^*)~,\\
  \Omega^\bullet_2(M,\sA)\ &=\ \Omega^2(M,\frg)\oplus\Omega^3(M,\frh)\oplus\Omega^4(M,\frg^*)~,\\
  \Omega^\bullet_3(M,\sA)\ &=\ \Omega^3(M,\frg)\oplus\Omega^4(M,\frh)\oplus\Omega^5(M,\frg^*)~,\\
  \Omega^\bullet_4(M,\sA)\ &=\ \Omega^4(M,\frg)\oplus\Omega^5(M,\frh)~,\\
  \Omega^\bullet_5(M,\sA)\ &=\ \Omega^5(M,\frg)~.
 \end{aligned}
\end{equation}
We can decompose a degree 1 element $\hat a\in \Omega^\bullet(M,\sA)$ as 
\begin{equation}
 \hat a\ :=\ A+B+C\ewith A\ \in\ \Omega^1(M,\frg)~,~B\ \in\ \Omega^2(M,\frh)~,~C\ \in\ \Omega^3(M,\frg^*)~.
\end{equation}
The homotopy Maurer--Cartan action~\eqref{eq:action_A_infty} for this degree~1 element reduces to 
\begin{equation}\label{eq:HCSAction}
\begin{aligned}
 S\ &=\ \int_M \Big\{\langle A,\dd C\rangle_\sA +\langle B,m_1(C)\rangle_\sA-\tfrac12 \langle B,\dd B\rangle_\sA+\langle A,m_2(A,C)\rangle_\sA\,+\\
 &\kern1.5cm+\langle A,m_2(B,B)\rangle_\sA+\langle A,m_3(A,A,B)\rangle_\sA+\tfrac15 \langle A,m_4(A,A,A,A)\rangle_\sA\Big\}~,
\end{aligned}
\end{equation}
where we simplified using cyclicity of the inner product on $ \Omega^\bullet(M,\sA)$ and the $m_i$ only act on the $\sA$-part of $A,B,$ and $C$. Note that varying the action creates terms of the form
\begin{equation}
m_i(a_1,\dots, a_i)+{\rm graded~cyclic}\ =\ \mu_i(a_1,\dots,a_i)~,~~~a_1,\dots,a_i\in \{A,B,C\}~,
\end{equation}
for the $\mu_i$ defined in~\eqref{eq:functor_A_to_L}. We can thus apply the functor from $A_\infty$-algebras to $L_\infty$-algebras and switch to the $L_\infty$-algebra picture. Correspondingly, the equations of motion then read as
\begin{equation}
\begin{aligned}
 \CF\ &:=\ \dd A+\tfrac 12 \mu_2(A,A)+\mu_1(B)\ =\ 0~,\\
 \CH\ &:=\ \dd B+\mu_2(A,B)-\tfrac{1}{3!}\mu_3(A,A,A)-\mu_1(C)\ =\ 0~,\\
 \CG\ &:=\ \dd C+\mu_2(A,C)+\tfrac12\mu_3(A,A,B)+\tfrac12 \mu_2(B,B)+\tfrac{1}{4!}\mu_4(A,A,A,A)\ =\ 0~,
\end{aligned}
\end{equation}
where $\CF$, $\CH$, and $\CG$ are the natural higher curvature forms which are elements of $\Omega^\bullet_2(M,\sA)$. The infinitesimal gauge transformations are parameterised by a degree~0 element $\lambda$ of $\Omega^\bullet(M,\sA)$, which we decompose into
\begin{equation}
 \lambda\ :=\ X+\Lambda+\Sigma\ewith X\ \in\ \CC^\infty(M,\frg)~,~\Lambda\ \in\ \Omega^1(M,\frh)~,~\Sigma\ \in\ \Omega^2(M,\frg^*)~.
\end{equation}
The infinitesimal gauge transformations are then read off~\eqref{eq:gauge_trafos_A_infty}:
\begin{equation}
\begin{aligned}
 \delta A\ &=\ \dd X-\mu_1(\Lambda)+\mu_2(A,X)~,\\
 \delta B\ &=\ \dd \Lambda+\mu_1(\Sigma)+\mu_2(B,X)+\mu_2(A,\Lambda)+\tfrac12\mu_3(A,A,X)~,\\
 \delta C\ &=\ \dd \Sigma+\mu_2(C,X)-\mu_2(B,\Lambda)+\mu_2(A,\Sigma)\,-\\
  &\kern2cm-\mu_3(A,B,X)-\tfrac12\mu_3(A,A,\Lambda)+\tfrac{1}{3!}\mu_4(A,A,A,X)~.
\end{aligned}
\end{equation}

\section{Yang--Mills theory from higher Chern--Simons theory}\label{sec:YMhCS}

\subsection{Ambitwistor space}

The {\it ambitwistor space} is the space of all light-rays in four-dimensional super Minkowski space. We shall be interested in $\CN=3$ super Minkowski space\footnote{Hence, only a subgroup of the R-symmetry group of MSYM theory will be manifest.}, which we complexify to $\FC^{4|12}$ for convenience. 

Making use of the identification $T\FC^4\cong S\otimes\tilde S$, where $S$ and $\tilde S$ are the bundles of chiral and anti-chiral spinors, we coordinatise $\FC^4$ in the standard fashion by $x^{\alpha\dot\alpha}$ with $\alpha,\beta,\ldots,\dot\alpha,\dot\beta,\ldots=1,2$. 
Spinorial indices are raised and lowered with the standard symplectic structures $\eps_{\alpha\beta}$ and $\eps_{\ald\bed}$ on $S$ and $\tilde S$, respectively. The Gra\ss mann-odd (fermionic) directions are coordinatised by $\theta^{i\alpha}$ and $\eta_i^{\dot\alpha}$ with $i,j,\ldots=1,\ldots,3$. Light-rays through a point $(x^{\alpha\dot\alpha}_0,\theta^{i\alpha}_0, \eta^{\dot\alpha}_{0\,i})\in \FC^{4|12}$ take the form
\begin{equation}\label{eq:superLightRay}
\begin{gathered}
x^{\alpha\dot\alpha}\ =\ x^{\alpha\dot\alpha}_0+t\mu^\alpha\lambda^{\dot\alpha} + t^i\mu^\alpha\eta_i^{\dot\alpha}-t_i\theta^{i\alpha}\lambda^{\dot\alpha}~,\\
\theta^{i\alpha}\ =\ \theta^{i\alpha}_0+t^i\mu^\alpha~,~~~
\eta^{\dot\alpha}_i\ =\ \eta^{\dot\alpha}_{0\,i}+t_i\lambda^{\dot\alpha}
\end{gathered}
\end{equation}
with $(t,t^i,t_i)\in\FC^{1|6}$, and they are thus parameterised by variables $(\mu_\alpha,\lambda_\ald)\in \PP^1\times \PP^1$. Together with the base point in $\FC^{4|12}$, they form the space $F^{6|12}:=\FC^{4|12}\times\PP^1\times\PP^1$. We can now factor out the dependence of the light-rays on the base point. Note that the vector fields generating translations along the light-rays are 
\begin{subequations}
\begin{equation}\label{eq:superVectorFieldsDF}
 V\ :=\ \mu^\alpha\lambda^{\dot\alpha}\partial_{\alpha\dot\alpha}~,~~~
 V_i\ :=\ \mu^\alpha D_{i\alpha}~,~~~
 V^i\ :=\ \lambda^{\dot\alpha} D_{\dot\alpha}^i~,
 \end{equation}
 where
 \begin{equation}
  D_{i\alpha}\ :=\ \partial_{i\alpha} + \eta^{\dot\alpha}_i\partial_{\alpha\dot\alpha}\eand
  D^i_{\dot\alpha}\ :=\ \partial^i_{\dot\alpha}+\theta^{i\alpha}\partial_{\alpha\dot\alpha}~.
 \end{equation}
 \end{subequations}

The space of orbits of $F^{6|12}$ under the translation group is then the ambitwistor space $L^{5|6}$, which is a rank $3|6$ holomorphic vector bundle over $\PP^1\times \PP^1$. This bundle is a subbundle of the rank $4|6$ holomorphic vector bundle $E:=\FC^{2|3}\otimes \big(\CO(1,0)\oplus \CO(0,1)\big)$ and given by the quadric
\begin{equation}
 z^\alpha\mu_\alpha-w^{\dot\alpha}\lambda_{\dot\alpha}+2\theta^i\eta_i\ =\ 0~,
\end{equation}
where $(z^\alpha,\eta_i)$ and $(w^\ald,\theta^i)$ are fibre coordinates on $\FC^{2|3}\otimes \CO(1,0)$ and $\FC^{2|3}\otimes \CO(0,1)$, respectively. In terms of these coordinates, the projection from $F^{6|12}$ onto the orbits reads as 
\begin{equation}\label{eq:superIncidenceRelations}
 z^\alpha\ =\ (x^{\alpha\dot\alpha}-\theta^{i\alpha}\eta_i^{\dot\alpha})\lambda_{\dot\alpha}~,~~
 \eta_i\ =\ \eta_i^{\dot\alpha}\lambda_{\dot\alpha}~,~~
 w^{\dot\alpha}\ =\ (x^{\alpha\dot\alpha}+\theta^{i\alpha}\eta_i^{\dot\alpha})\mu_\alpha~,~~
 \theta^i\ =\ \theta^{i\alpha}\mu_\alpha~,
\end{equation}
and we obtain the double fibration
\begin{equation}\label{eq:superDoubleFibration}
 \begin{picture}(50,40)
  \put(0.0,0.0){\makebox(0,0)[c]{$L^{5|6}$}}
  \put(64.0,0.0){\makebox(0,0)[c]{$\FC^{4|12}$}}
  \put(34.0,33.0){\makebox(0,0)[c]{$F^{6|12}$}}
  \put(7.0,18.0){\makebox(0,0)[c]{$\pi_1$}}
  \put(55.0,18.0){\makebox(0,0)[c]{$\pi_2$}}
  \put(25.0,25.0){\vector(-1,-1){18}}
  \put(37.0,25.0){\vector(1,-1){18}}
 \end{picture}
\end{equation}
with $\pi_2$ trivial. 

For our purposes, it is important to note that $L^{5|6}$ is a {\it formal} Calabi--Yau supermanifold\footnote{Here, `formal' refers to the fact that for supermanifolds the vanishing of the first Chern class does not necessarily guarantee the Ricci flatness~\cite{Rocek:2004bi}.} with global holomorphic measure
\begin{equation}\label{eq:HolomorphicMeasure}
 \Omega^{5|6,0}\ :=\ \oint_\gamma \frac{\dd z^\alpha \wedge \dd z_\alpha\wedge\dd \lambda^{\dot\alpha}\lambda_{\dot\alpha}\wedge\dd w^{\dot\beta}\wedge\dd w_{\dot\beta}\wedge\dd \mu^\beta\mu_\beta\otimes \dd\eta_1\dd\eta_2\dd\eta_3\dd\theta^1\dd\theta^2\dd\theta^3}{z^\alpha\mu_\alpha-w^{\dot\alpha}\lambda_{\dot\alpha}+2\theta^i\eta_i}~.
\end{equation}
Here, $\gamma$ is a contour encircling $L^{5|6}\hookrightarrow E$ while $\dd\eta_1\dd\eta_2\dd\eta_3\dd\theta^1\dd\theta^2\dd\theta^3$ is to be understood as an integral form in the sense of Berezin.

\subsection{Higher ambitwistor space action}

By means of the double fibration~\eqref{eq:superDoubleFibration} and incidence relations~\eqref{eq:superIncidenceRelations}, points $p:=(x,\theta,\eta)\in \FC^{4|12}$ correspond to (holomorphic) embeddings of $\hat p:=\pi_1(\pi^{-1}_2(x,\theta,\eta))\cong \PP^1\times\PP^1\hookrightarrow L^{5|6}$. We shall be interested in smoothly trivial holomorphic principal $\sG$-bundles over  $L^{5|6}$ that are holomorphically trivial when restricted to $\hat p$ for any $p\in \FC^{4|12}$. Following Manin, we shall refer to such bundles as {\it $\FC^{4|12}$-trivial}. Due to~\cite{Witten:1978xx,Isenberg:1978kk,Eastwood:1987aa} we now have the following result.

{\theorem\label{thm:PenroseWard}
There is a bijection between gauge equivalence classes of complex holomorphic solutions to the MSYM equations on $\FC^4$ with gauge group $\sG$ and equivalence classes of $\FC^{4|12}$-trivial holomorphic principal $\sG$-bundles over $L^{5|6}$.
}

\vspace{10pt}
\noindent
Roughly speaking, the holomorphic triviality on all $\hat p$ guarantees that the pullback of this bundle to the correspondence space $F^{6|12}$ is holomorphically trivial on all of $F^{6|12}$. This, in turn, implies that the transition functions of this bundle can be split holomorphically. Furthermore, these transition functions are annihilated by the vector fields~\eqref{eq:superVectorFieldsDF}, and this can be used to find an auxiliary linear system of partial differential equations on $\FC^{4|12}$. Following~\cite{Harnad:1984vk,Harnad:1985bc}, the compatibility conditions of this linear system are equivalent to the MSYM equations equations with gauge group $\sG$ on $\FC^4$ (or on $\FR^{1,3}$ after choosing suitable reality conditions). For a review on the details of this construction, we refer e.g.~to~\cite{Popov:2004rb} or~\cite{Manin:1988ds}.

An alternative way of describing holomorphic principal $\sG$-bundles is in terms of the Dolbeault picture. Consider a complex manifold $X$  with an open cover $\{U_a\}$ and equipped with a complex principal $\sG$-bundle $P$ described by the transition functions $\{g_{ab}:U_a\cap U_b\to\sG\}$. A $(0,1)$-connection on $P$ is a collection $\{A_a^{0,1}\in\Omega^{0,1}(U_a,\sLie(\sG))\}$ of locally defined $(0,1)$-forms on $X$ with values in $\sLie(\sG)$ which obey
\begin{equation}\label{eq:GT01Form}
 A_b^{0,1}\ =\ g_{ab}^{-1}A_a^{0,1}g_{ab}+g_{ab}^{-1}\bar\partial g_{ab}\quad{\rm on}\quad U_a\cap U_b~,
\end{equation}
where $\bar\partial$ is the anti-holomorphic exterior derivative on $X$. The associated curvature $(0,2)$-form $\{F_a^{0,2}\in\Omega^{0,2}(U_a,\sLie(\sG))\}$ is defined by
\begin{equation}
 F_a^{0,2}\ :=\ \bar\partial A_a^{0,1}+\tfrac12[A_a^{0,1}, A_a^{0,1}]~,
\end{equation}
and we have
\begin{equation}
 F_b^{0,2}\ =\ g_{ab}^{-1}F_a^{0,2}g_{ab}\quad{\rm on}\quad U_a\cap U_b~.
\end{equation}
Automorphisms on $P$ are locally given by smooth functions $g_a:U_a\to\sG$ which obey\footnote{These transformations should not be confused with usual coboundary transformations (i.e.~gauge transformations) which are mediated by unconstrained functions $g_a:U_a\to\sG$ with $g_{ab}\mapsto\tilde g_{ab}:=g_{a}g_{ab}g_b^{-1}$. Note that the set of all automorphisms forms a group and it can be identified with the space of global sections of the associated bundle $P\times_{\sAd} \sG$, where $\sAd$ is the adjoint map.}
\begin{equation}\label{eq:AutAct}
 g_b\ =\ g_{ab}^{-1} g_a g_{ab}\quad \Longleftrightarrow\quad g_{ab}\ =\ g_{a}g_{ab}g_b^{-1} \quad{\rm on}\quad U_a\cap U_b~.
\end{equation}
They act on $A_a^{0,1}$ and $F_a^{0,1}$ as 
\begin{equation}\label{eq:Aut01Form}
 A_a^{0,1}\ \mapsto\ g_{a}^{-1}A_a^{0,1}g_{a}+g_{a}^{-1}\bar\partial g_{a}\eand
 F_a^{0,2}\ \mapsto\ g_{a}^{-1}F_a^{0,2}g_{a}~.
\end{equation}
The bundle $P$ is said to be holomorphic provided that
\begin{equation}\label{eq:EOMHCS}
  F_a^{0,2}\ =\ 0\quad{\rm on}\quad U_a~.
\end{equation}
This equation is generally solved by
\begin{equation}
 A_a^{0,1}\ =\ \psi^{-1}_a\bar\partial \psi_a
\end{equation} 
for smooth maps $\psi_a:U_a\to\sG$, and, upon substituting this expression into~\eqref{eq:GT01Form}, we find
\begin{equation}
 \check g_{ab}\ :=\ \psi_ag_{ab}\psi_b^{-1}\ewith \bar\partial \check g_{ab}\ =\ 0~.
\end{equation}
Thus, we have obtained a holomorphic principal $\sG$-bundle $\check P$ with holomorphic transition functions $\{\check g_{ab}:U_a\cap U_b\to\sG\}$ that is smoothly equivalent to $P$. Furthermore, automorphisms act on the $\psi_a$ as $\psi_a\mapsto \psi_a g_a$, and, consequently, they leave the transition functions $\{\check g_{ab}:U_a\cap U_b\to\sG\}$ invariant. 

Hence, there is a bijection between the set of solutions to~\eqref{eq:EOMHCS} on $P$ modulo the action of the automorphism transformations~\eqref{eq:Aut01Form} and the set of all holomorphic principal $\sG$-bundles $\check P$ smoothly equivalent to $P$. This is known as the {\it \v Cech--Dolbeault correspondence}. See \cite{Popov:1998pc,Popov:1998fb,Wolf:2006me,Rink:2012aa} for detailed explanations.

We may thus shift the starting point of Theorem~\ref{thm:PenroseWard} to equation~\eqref{eq:EOMHCS}. That is, solving \eqref{eq:EOMHCS} in the case of a smoothly trivial bundle on $L^{5|6}$ is equivalent to solving the MSYM equations on $\FC^4$ under the assumption that there is a gauge in which the $(0,1)$-forms $\{A_a^{0,1|0}\}$ have no anti-holomorphic fermionic directions, depend holomorphically on the fermionic coordinates~\cite{Witten:2003nn} and are zero when restricted to $\hat p$ for any $p\in \FC^{4|12}$. While the former requirement on the gauge is not really a restriction since the fermionic fibres are contractible, the latter requirement coincides with the assumption of $\FC^{4|12}$-triviality. Here and in the following, the superscript `$-|0$' indicates that the differential forms under consideration have no anti-holomorphic fermionic directions and depend holomorphically on the fermionic coordinates. 

If $X$ is a three-dimensional Calabi--Yau manifold, then there is a globally defined no-where vanishing holomorphic volume form $\Omega^{3,0}$, and for smoothly trivial complex principal bundles~\eqref{eq:EOMHCS} follows from the action of {\it holomorphic Chern--Simons theory}~\cite{Witten:9207094}
\begin{equation}
 S\ =\ \int_X\Omega^{3,0}\wedge{\rm tr} \left\{A^{0,1}\wedge\bar\partial A^{0,1}+\tfrac23 A^{0,1}\wedge A^{0,1}\wedge A^{0,1}\right\},
\end{equation}
which we also obtain from Section~\ref{ssec:HigherChanPaton} for $d=3$. This action is well-defined since under the above assumptions, $A^{0,1}$ is globally defined, that is, $A^{0,1}|_{U_a}=A^{0,1}_a$ and $A^{0,1}_a=A^{0,1}_b$. Unfortunately, in the case of $L^{5|6}$ it is not directly possible to use this action\footnote{In~\cite{Mason:2005kn}, an ambidextrous Chern--Simons action was formulated that requires the selection of a certain codimension 2 Cauchy--Riemann submanifold (in ambitwistor space) which is possible for a real structure on $\FC^4$ associated with Euclidean signature (but not Minkowskian signature).}  since, despite admitting a globally defined no-where vanishing holomorphic volume $\Omega^{5|6,0}$ given in~\eqref{eq:HolomorphicMeasure}, the body of the supermanifold $L^{5|6}$ is five-dimensional. 

However, a natural action for $L^{5|6}$ is the higher Chern--Simons action~\eqref{eq:HCSAction} when reformulated for smoothly trivial complex principal 3-bundles. The kinematical data including a connective structure on principal 3-bundles bundles was described in detail in~\cite{Saemann:2013pca,Jurco:2016qwv}. The most general case, involving Lie quasi 3-groups $\CCG$, can be derived from the formalisms of~\cite{Fiorenza:2010mh} or~\cite{Jurco:2016qwv}. Such a quasi-group $\CCG$ comes with an associated 3-term $L_\infty$-algebra $\sL:=\sLie(\CCG)$ with $\sL:=\sL_{-2}\oplus \sL_{-1}\oplus \sL_{0}$ \cite{Severa:2006aa,Jurco:2016qwv}. Consequently, a degree~1 element $\hat a$ in $\Omega^{0,\bullet|0}(L^{5|6},\sL)$ is written as $\hat a=A^{0,1|0}+B^{0,2|0}+C^{0,3|0}$ with 
\begin{equation}\label{eq:HoloConnectiveStructure}
 A^{0,1|0}\in \Omega^{0,1}(L^{5|6},\sL_0)~,~~~B^{0,2|0}\in \Omega^{0,2}(L^{5|6},\sL_{-1})~,\eand C^{0,3|0}\in \Omega^{0,3}(L^{5|6},\sL_{-2})~.
\end{equation}
The homotopy Maurer--Cartan action \eqref{eq:action_A_infty} (see also \eqref{eq:cyclic_naive}) then takes the form
\begin{equation}\label{eq:HHCSAction}
\begin{aligned}
 \kern-5pt S\ &=\ \int_{L^{5|6}} \Omega^{5|6,0}\wedge\Big\{\langle A^{0,1|0},\bar\partial C^{0,3|0}\rangle_\sL+\langle B^{0,2|0},\mu_1(C^{0,3|0})\rangle_\sL-
 \tfrac12\langle B^{0,2|0},\bar\partial B^{0,2|0}\rangle_\sL\,+\\
 &\kern3cm+\tfrac{1}{2} \langle A^{0,1|0},\mu_2(A^{0,1|0},C^{0,3|0})\rangle_\sL+\tfrac{1}{2}\langle A^{0,1|0},\mu_2(B^{0,2|0},B^{0,2|0})\rangle_\sL\,+\\
 &\kern4cm+\tfrac{1}{3!}\langle A^{0,1|0},\mu_3(A^{0,1|0},A^{0,1|0},B^{0,2|0})\rangle_\sL\,+\\
 &\kern5cm+\tfrac{1}{5!}\langle A^{0,1|0},\mu_4(A^{0,1|0},A^{0,1|0},A^{0,1|0},A^{0,1|0})\rangle_\sL\Big\}~,
\end{aligned}
\end{equation}
where the $\mu_i$ are the higher products in $\sL$. The corresponding equations of motion on the coordinate patch $U_a$ read as
\begin{equation}\label{eq:HHCSEoM}
\begin{aligned}
 \CF^{0,2|0}_a\ &:=\ \bar\partial A^{0,1|0}_a+\tfrac{1}{2} \mu_2(A^{0,1|0}_a,A^{0,1|0}_a)+\mu_1(B^{0,2|0}_a)\ =\ 0~,\\
 \CH^{0,3|0}_a\ &:=\ \bar\partial B^{0,2|0}_a+\mu_2(A^{0,1|0}_a,B^{0,2|0}_a)-\tfrac{1}{3!}\mu_3(A^{0,1|0}_a,A^{0,1|0}_a,A^{0,1|0}_a)-\mu_1(C^{0,3|0}_a)\ =\ 0~,\\
 \CG^{0,4|0}_a\ &:=\ \bar\partial C^{0,3|0}_a+\mu_2(A^{0,1|0}_a,C^{0,3|0}_a)+ \tfrac{1}{2} \mu_2(B^{0,2|0}_a,B^{0,2|0}_a)\,+ \\
  &\kern1cm+ \tfrac{1}{2} \mu_3(A^{0,1|0}_a,A^{0,1|0}_a,B^{0,2|0}_a) + \tfrac{1}{4!}\mu_4(A^{0,1|0}_a,A^{0,1|0}_a,A^{0,1|0}_a,A^{0,1|0}_a)\ =\ 0~.
\end{aligned}
\end{equation}
For the action~\eqref{eq:HHCSAction} to make sense\footnote{Obviously, this action makes always sense on an ordinary five-dimensional Calabi--Yau manifold. Furthermore, the equations of motion~\eqref{eq:HHCSEoM} make also sense for smoothly non-trivial bundles.}, we follow Witten~\cite{Witten:2003nn} and assume that the connective structure $\{A^{0,1|0}_a,B^{0,2|0}_a,C^{0,3|0}_a\}$ has no anti-holomorphic fermionic directions and depends holomorphically on the fermionic coordinates. We shall refer to~\eqref{eq:HHCSAction} as the {\it holomorphic higher Chern--Simons action}. 

It is important to stress that the gauge potential $C^{0,3|0}_a$ should not be understood as a simple Lagrange multiplier because it is part of a higher connective structure and obeys a different transformation law from a Lagrange multiplier in the adjoint representation of some ordinary Lie group. Only if we forced $B^{0,2|0}_a=0$ and chose a 3-term $L_\infty$-algebra arising from ordinary Chan--Paton factors as discussed in Section~\ref{ssec:ordinary_Chan_Paton}, then we could relate the equations of motion of the field $C^{0,3|0}_a$ to those of a Lagrange multiplier. This, however, corresponds to a vast restriction of the symmetries of the action~\eqref{eq:HHCSAction}. The latter is a more natural starting point for studying e.g.\ scattering amplitudes from a twistor string theoretic perspective as will become clear shortly.

Before moving on, we note that the \v Cech--Dolbeault correspondence for ordinary principal bundles we discussed previously generalises to higher principal bundles. The details of a proof are found in~\cite{Saemann:2011nb,Saemann:2012uq,Saemann:2013pca,Jurco:2014mva,Demessie:2014ewa,Jurco:2016qwv}.\footnote{These papers prove a version of the \v Cech--Dolbeault correspondence for a relative exterior derivative along a fibration. However, the arguments are completely analogous for the $\bar\partial$-operator on a complex manifold $X$. This is the case since the higher cohomology groups with values in the sheaf of smooth differential $(0,q)$-forms $\Omega^{0,q}$, which are needed to verify the correspondence for relative differential forms, vanish trivially as this is a fine sheaf when working in the smooth category.}  Consequently, we may describe holomorphic principal 3-bundles over a complex manifold either by complex principal 3-bundles equipped with a connective structure of the form \eqref{eq:HoloConnectiveStructure} and subject to~\eqref{eq:HHCSEoM} (the Dolbeault picture) or, equivalently, by holomorphic transition functions (the \v Cech picture).

We now wish to interpret equations~\eqref{eq:HHCSEoM} further and connect them to Theorem~\ref{thm:PenroseWard}. To do so, we shall first assume that there is a gauge for the gauge potentials $\{A^{0,1|0}_a,B^{0,2|0}_a,\newline C^{0,3|0}_a\}$ in which their restriction to $\hat p  \hookrightarrow L^{5|6}$ for any $p\in \FC^{4|12}$ vanishes. By the \v Cech--Dolbeault correspondence, this will yield a $\FC^{4|12}$-trivial holomorphic principal 3-bundle (i.e.~smoothly trivial on $L^{5|6}$ and holomorphically trivial on all $\hat p \hookrightarrow L^{5|6}$).

Next, we note that our constructions of principal 3-bundles and connections via the \v Cech--Dolbeault correspondence are now transparent to categorical and higher categorical equivalence. In particular, any 3-isomorphism between any two Lie 3-groups $\CCG$ and $\CCG'$ induces a 3-isomorphism between equivalence classes of principal $\CCG$-bundles and principal $\CCG'$-bundles. This, in turn, induces a 3-isomorphism between the equivalence classes of solutions to the higher Chern--Simons equations \eqref{eq:HHCSEoM} for the associated $3$-term $L_\infty$-algebras $\sL:=\sLie(\CCG)$ and $\sL':=\sLie(\CCG')$. Hence, without loss of generality, we may either work with $\sL$ or $\sL'$ to interpret the solutions to \eqref{eq:HHCSEoM}.

Furthermore, we recall {\em Kadeishvili's theorem} \cite{Kadeishvili:0504437} for $A_\infty$-algebras and its induced version on $L_\infty$-algebras. This theorem states that any $A_\infty$-algebra is categorically equivalent to a {\it minimal $A_\infty$-algebra} which is an $A_\infty$-algebra with $m^{\rm min}_1=0$. Consequently, on-shell, we can work with a categorically equivalent $3$-term $L_\infty$-algebra that has $\mu^{\rm min}_1=0$ so that the first equation of \eqref{eq:HHCSEoM} simplifies to 
\begin{equation}
 \CF^{0,2|0}_a\ =\ \bar\partial A^{0,1|0}_a+\tfrac{1}{2} \mu^{\rm min}_2(A^{0,1|0}_a,A^{0,1|0}_a)\ =\ 0~.
\end{equation}
As we have assumed that there is a gauge for $\{A^{0,1|0}_a,B^{0,2|0}_a,C^{0,3|0}_a\}$ in which their restriction to $\hat p  \hookrightarrow L^{5|6}$ for any $p\in \FC^{4|12}$ vanishes, this is equivalent to the MSYM equations on $\FC^4$ by virtue of Theorem~\ref{thm:PenroseWard}. It remains to clarify the content of the remaining potentials $B^{0,2|0}_a$ and $C^{0,3|0}_a$ in this context. They satisfy equations in the {\it background} of $A^{0,1|0}_a$ and hence, they decouple from the equation for $A^{0,1|0}_a$.  In addition, they are mapped to fields, collectively denoted by $\{\Phi\}$, on complexified Minkowski space $\FC^4$ via the Penrose--Ward transform. Since all our constructions are manifestly covariant under $\CN=3$ superconformal symmetry as well as under parity transformations, the resulting equations of motion for the fields $\{\Phi\}$ on $\FC^4$ have to be $\CN=4$ supersymmetric. However, the only such equations are the MSYM equations. Consequently, the fields $\{\Phi\}$ would have to transform either trivially under $\CN=4$ supersymmetry or decompose into $\CN=4$ vector supermultiplets. Both cases are excluded by gauge covariance with respect to the gauge potential contained in the vector supermultiplet arising from $A^{0,1|0}_a$. Therefore, the gauge potentials $B^{0,3|0}_a$ and $C^{0,3|0}_a$ cannot contain any additional on-shell degrees of freedom. We are thus led to conclude that the equations of motion~\eqref{eq:HHCSEoM} for the holomorphic higher Chern--Simons action~\eqref{eq:HHCSAction} on $L^{5|6}$ are equivalent to the equations of motion of ordinary holomorphic Chern--Simons theory. 

The summary of our findings is now the following statement: at the classical level, holomorphic higher Chern--Simons theory on $L^{5|6}$ described by \eqref{eq:HHCSAction} and \eqref{eq:HHCSEoM} is equivalent to MSYM theory on $\FC^4$. After choosing suitable reality conditions, this equivalence extends to MSYM theory on $\FR^{1,3}$.

\section{Comments on higher dimensions}

In this last section, let us briefly comment on higher dimensional twistor spaces. There are in particular two spaces that come to mind: 
\begin{enumerate}[i)]
 \item The ambitwistor space $L^{9|8}$ was introduced in~\cite{Saemann:2012rr}. It is the moduli space of light-rays in six-dimensional Minkowski superspace and it can be used to describe solutions to the MSYM equations in six dimensions.
 \item The twistor space $P^{6|4}$ as defined in~\cite{Saemann:2012uq}, see also~\cite{Mason:2012va}, is the moduli space of isotropic  planes in six-dimensional Minkowski chiral superspace and provides a twistorial description of solutions to $\CN=(2,0)$ superconformal field equations in six dimensions.
\end{enumerate}
There is, however, an issue with both of these spaces. Neither of them is a Calabi--Yau supermanifold, not even in the weak sense necessary here. That is, they do not carry a global holomorphic measure which would be required for writing down holomorphic higher Chern--Simons action functionals. Nevertheless, we can still consider the equations of motion of appropriate holomorphic higher Chern--Simons theories, which do cover the relevant higher bundles over the twistor space to capture both maximally supersymmetric field theories in six dimensions.

The ambitwistor space $L^{9|8}$ has a nine-dimensional body, and correspondingly it makes sense to consider the equations of motion of holomorphic higher Chern--Simons theory based on Lie $n$-algebras with $n\leq 7$. Just as in the case of the ambitwistor space $L^{5|6}$, however, the bosonic fibres of the correspondence space over twistor space are one-dimensional. Therefore we expect that the extension from Lie algebras to higher Lie algebras does not yield any further information at the level of equations of motion.

The twistor space $P^{6|4}$ has a six-dimensional body, and we can thus consider the equations of motion of holomorphic higher Chern--Simons theory based on Lie $n$-algebras with $n\leq 4$. The bosonic fibres of the correspondence space over twistor space are now three-dimensional, and this suggests that the extension from $n=3$ to $n=4$ turns out to be trivial at the level of equations of motion.

Note that solutions to the holomorphic higher Chern--Simons theories on $L^{5|6}$ give rise to solutions on both $L^{9|8}$ and $P^{6|4}$. In the first case, this involves a simple pullback along the projection $L^{9|8}\rightarrow L^{5|6}$, which is detailed in~\cite{Saemann:2012rr}. The second case is more involved, since the form degree will jump: note that there is a projection from a subspace of $P^{6|4}$ to $L^{5|6}$. By looking at the underlying cocycles in the obvious covering, it is not hard to see (at least in the Abelian case) that the pullback of a holomorphic principal 1-bundle over $L^{5|6}$ to that subspace encodes a principal 2-bundles over $P^{6|4}$. Analogously, a holomorphic 1-form on $L^{5|6}$ will give rise to a holomorphic 2-form on $P^{6|4}$.

\acknowledgements

We would like to thank Jan Gutowski, Branislav Jur{\v c}o, Alessandro Torrielli, and, in particular, Lotte Hollands for useful discussions. We are also grateful to Michael Eastwood for very helpful correspondence and Lorenzo Raspollini for comments on the first version. C.S.~was supported in part by the STFC Consolidated Grant ST/L000334/1 {\it Particle Theory at the Higgs Centre}. M.W.~was supported in part by the STFC Consolidated Grant ST/L000490/1 {\it Fundamental Implications of Fields, Strings, and Gravity}.

\datamanagement

No additional research data beyond the data presented and cited in this work are needed to validate the research findings in this work.

\bibliographystyle{latexeu}

\bibliography{littleone}



\end{document}